\documentstyle[11pt,aaspp4,psfig]{article}

\begin{document}

\slugcomment{ApJ (Pt.1). Accepted}

\title{An inverse Compton scattering (ICS) model of pulsar emission: \\
III. polarization}

\author{
R. X. Xu\altaffilmark{1,2,3},
J. F. Liu\altaffilmark{2,3},
J. L. Han\altaffilmark{1,3},
G. J. Qiao\altaffilmark{4,2,3}
}

\altaffiltext{1}{
  Beijing Astronomical Observatory and National Astronomical
  Observatories, Chinese Academy of Sciences (CAS), Jia 20 Da-Tun Road,
  Chao-Yang District, Beijing 100012, China}
\altaffiltext{2}{
  Astrophysics Devision, Geophysics Department, Peking University (PKU),
  Beijing 100871, China}
\altaffiltext{3}{
  Beijing Astrophysical Center, CAS-PKU, Beijing 100871, China}
\altaffiltext{4}{CCAST (World Laboratory) P.O.Box 8730, Beijing 100080, China}

\begin{abstract}

Qiao and his collaborators recently proposed an inverse Compton
Scattering (ICS) model to explain radio emission of pulsars. In
this paper, we investigate the polarization properties of pulsar
emission in the model. First of all, using the lower frequency
approximation, we derived the analytical amplitude of inverse Compton
scattered wave of a single electron in strong magnetic field. We
found that the out-going radio emission of a single relativistic
electron scattering off the ``low frequency waves'' produced by
gap-sparking should be linearly polarized and have no circular
polarization at all. However, considering the coherency of
the emission from a bunch of electrons, we found that the out-going
radiation from the inner part of emission beam, i.e., that from the
lower emission altitudes, prefers to have circular polarization.
Computer simulations show that the polarization properties, such as
the sense reversal of circular polarization near the pulse center,
S-shape of position angle swing of the linear polarization, strong
linear polarization in conal components, can be reproduced in the
ICS model.

\noindent
{\it Subject Headings}:  {polarization --- Pulsars: general --- Radiation
	  mechanisms: non-thermal}

\end{abstract}

\section{Introduction}

The outstanding polarization properties of radio pulsars are
keys to understand the magnetospheric structures of the neutron
stars and the unknown emission mechanisms. Not only linear
but also circular polarization has been detected from most
pulsars. Soon after the discovery of pulsars,
\markcite{rc69}Radhakrishnan \& Cook (1969) proposed the
rotation vector model to interpret the rotating position
angle of linear polarization, which has been widely accepted
for practical reasons, regardless of the emission
mechanism. Obviously, the linear polarization is probably
related to the structure of the strong dipole field above
magnetic poles. Large amount of polarization data have been
accumulated (e.g. Gould \& Lyne 1998; \markcite{gl98}
\markcite{mhq98} Manchester, Han \& Qiao 1998;
\markcite{rsw89} Rankin, Stinebring \& Weisberg 1989;
Weisberg et al. 1999\markcite{weiet99}).
However, many observed polarization features can not be
well explained, for example, the orthogonal position angles of
linear polarization observed from individual pulses
(e.g. Stinebring et
al. 1984a,b\markcite{sti84a}\markcite{sti84b}), diverse
circular polarization \markcite{hmxq98} (e.g. Han et
al. 1998), the different polarization characteristics of
core and conal emission (Rankin 1993;
\markcite{ran93}\markcite{lm88}Lyne \& Manchester
1988). Though there were many theoretical efforts (e.g. Gil
\& Snakowski 1990\markcite{gs90}), no consensus has ever
reached on how pulsar polarization was generated
(Radhakrishnan \markcite{rad92}1992; Melrose 1995\markcite{mel95}).
For example, the
observed circular polarization in pulsar radio emission
could either be converted from the linear
polarization in the pulsar magnetosphere via the propagation
effect (Melrose 1979\markcite{mel79},
Cheng \& Ruderman\markcite{cr79} 1979; von\markcite{vl99}
Hoensbroech \& Lesch 1999) or plasma process (e.g. Kazbegi
et al.  \markcite{kmm91}1991), or origin from emission
process intrinsically (Radhakrishnan \& Rankin
\markcite{rr90}1990; Michel 1987\markcite{mic87}; Gangadhara
1997\markcite{gan97}).

Recently, Qiao and his collaborators proposed an inverse
Compton scattering (ICS) model of pulsar emission (Qiao
1988\markcite{qiao88}; Qiao \& Lin\markcite{ql98} 1998; Qiao et al. 
1999). In the model, radio emission is the result of the secondary
particles scattering off ``low-frequency waves''. The waves
are assumed to be produced by either the breaking down of the vacuum
polar gap, or other sorts of short time oscillations or
micro-instabilities (Bj\"ornsson 1996) \markcite{bjo96}.
The gap-sparking may exist above polar cap, as shown
by new observations by Deshpande \& Rankin (1999)\markcite{dr99}
and by Vivekanand \& Joshi (1999)\markcite{vj99}. If so, there
must be low-frequency waves emitted by such sparking. The time
scale of a sparking at a certain point is about $10^{-5}$ s or
even shorter, the out-flowing plasma produced by the cascades of
the sparking should form miniflux tubes with a radius of $\sim 10$ m
and a length of $\sim 100$ m, and hence be highly inhomogenous in space.
The plasma density between the tubes should be sufficiently low
for radio wave to propagate, as if in a vacuum. Therefore, we assume
that the pulsar magnetosphere is transparent for radio emission.
That is to say, electromagnetic waves of observed emission can pass through the
magnetosphere freely,  and the so-called ``low-frequency
wave'' can propagate through the emission region above the
polar gap (see Qiao \& Lin 1998\markcite{ql98}).
The success of the ICS model\markcite{ql98}
(Qiao \& Lin 1998, Qiao et al. 1999)\markcite{qlzh99}
is that the core and the conal emission
beams can be naturally explained in this model, while the
core component comes from the nearest region to neutron-star
surface, the inner cone is farther, and the out cone is
farthest. Because of different heights of these emission
components, the retardation and aberration effects cause the
components apparently shift spatially, and could produce the
position angle jumps in the integrated pulse profiles in some cases
(e.g. Xu et al. 1997\markcite{xqh97}).

In this paper, we investigate the polarization properties of
the ICS model for radio pulsars. First of all, we presented
the polarization characteristics of scattered radio emission
from a single particle in Sect.~2, half from theoretical
work, half from numerical calculations. In the low frequency
limit, deduction of the amplitude of outgoing radio waves
from the ICS process in classical electrodynamics (CED) was
presented in Appendix. In Sect.~3, we showed that the
circular polarization can be the result of coherency of the
outgoing waves. Considering the geometry above the polar cap
regions, we numerically simulated the ICS process, and found
that the circular polarization with possible sense reversal
is preferably found from the emission in the beam center,
while stronger linear polarization preferably from the outer
part of the beam. Some related issues will be discussed in
Sect.~4.

\section{Polarization features of the out-going scattered waves from
a single electron}

The polarization features of scattered emission by
relativistic electrons in the {\it strong} magnetic field
has not been discussed in literature.  In this section we
are concerned with the general polarization properties of the
scattered photons by a single relativistic electron onto the
lower frequency waves. Previously, the quantum
electrodynamics (QED) results of Compton scattering cross
sections were first derived by Herold (1979)
\markcite{her79} for electrons with ground initial and final
states. Then, the QED cross sections for various initial and
final electron states were calculated by Melrose \& Parle
(1983) and \markcite{mp83} Daugherty \& Harding
(1986)\markcite{dh86}. Finally, Bussard, Alexander \&
Mesaros (1986)\markcite{bam86} presented the QED results for
electrons with arbitrary initial and final states. Xia et
al. (1985) \markcite{xia85} calculated for the first time
the differential cross section of {\it inverse} Compton
scattering, but no polarization features were discussed.

Based on the results of cross sections of various
polarization vectors derived by Herold
(1979)\markcite{her79}, we used the Lorentz transformation
in the magnetic field direction to calculate the
differential cross section of inverse Compton scattering of
orthogonal polarization pairs, an approach similar to
that in Xia et al. (1985)\markcite {xia85}.  Hereafter we
denote the angular frequency of an incident photon as
$\omega_{\rm in}$, the strength of magnetic field as $B$,
and its cyclotron frequency as $\omega_{\rm c}={eB\over mc}$,
and the angular frequency of the scattered wave or photon as
$\omega_{\rm out}$. The geometry of the ICS process is shown
in Fig.~1. The relativistic electron is moving along the
field line (on the $z$ axis) and has a Lorentz factor
$\gamma$ (and $\beta^2=1-1/\gamma^2$ in the equations) which
we suppose to be about $\sim 10^2$ to $\sim 10^3$ in the
pulsar magnetosphere. From the kinematic restrictions for the
scattering process in strong magnetic fields, i.e. the
momentum conservation in the magnetic field direction and
the energy conservation of the process, the frequency of the
scattered photons is (note that ``$\sin\theta$'' in
Eq.(15c) of Xia et al. 1985\markcite{xia85} should read
as``$\sin^2\theta$'')
$$
\omega_{\rm out} = 
\left\{    \begin{array}{ll}
\omega_{\rm in} {\gamma(1+\beta)\over 2}
	{2\gamma m c^2(1-\beta\cos\theta_{\rm in})
            + \hbar \omega_{\rm in} \sin^2\theta_{\rm in} 
   \over  mc^2+\gamma\hbar\omega_{\rm in}(1+\beta)(1 - \cos\theta_{\rm in})},
& \theta_{\rm out}=0 \\
\omega_{\rm in} {2 \gamma m c^2(1-\beta\cos\theta_{\rm in}) +
	\hbar \omega_{\rm in}
\sin^2\theta_{\rm in} 
\over 2[\gamma(1+\beta)mc^2+\hbar\omega_{\rm in} (1 + \cos\theta_{\rm in})]},
& \theta_{\rm out}=\pi \\
{f( 1- \sqrt{1-g})
        \over \hbar \sin^2\theta_{\rm out}},
& \theta_{\rm out}\neq 0,\pi
\end{array}     \right.
\eqno(1)
$$
where
$ f=\gamma(1-\beta\cos\theta_{\rm out})mc^2 +
    (1-\cos\theta_{\rm in}\cos\theta_{\rm out})\hbar\omega_{\rm in}$ and 
$ g= \hbar\omega_{\rm in}\sin^2\theta_{\rm out}
[2\gamma(1-\beta\cos\theta_{\rm in})mc^2
	+\sin^2\theta_{\rm in}\hbar\omega_{\rm in}]/f^2.
$
When $\hbar \omega_{\rm in} \ll m_{\rm e} c^2 $, all above three equations
can be simplified as being
$$
\omega_{\rm out} = \omega_{\rm in} {1 - \beta \cos\theta_{\rm in} \over
                    1 - \beta\cos\theta_{\rm out}}.    \eqno(2)
$$

According to the definition given by Saikia (1988)\markcite{saik88}, we
derived the Stokes parameters (similar to Eq.(6) below) of scattered
wave with the geometry plotted in Fig.~1. We numerically calculated
the polarization properties, using Herold's results, for radio frequency
regime and high energy bands. We found that the polarization properties
of the scattered photons are different.

(1). When $\omega_{\rm in}\ll \omega_c$ and both $\omega_{\rm
in}$ and $\omega_{\rm out}$ are in radio band, the scattered
photons are {\it completely} linearly polarized, and its
polarization position angle (PA) is {\it definitely} in the
co-plane of the out-going photon direction and the magnetic
field (see geometry in Fig.~1), in spite of the polarization
state of incident photons with one exception for the very specific
case in which the incident photon is absolutely linear polarized
{\it and} its polarization vector is exactly perpendicular
to the magnetic field line.

We derived the amplitude of the scattered waves from
a single electron with the classical electrodynamics (CED),
which we will use in next sections for emission coherence.
The electron can be considered as a particle since the
magnetic field $B\ll B_q$, here $B_q$ is the critical
magnetic field $B_q = (m^2 c^3)/(e \hbar) = 4.413 \times
10^{13}$ G, and the incident and the scattered photons can
be considered as waves since the frequency of the incident
wave in the electron-rest-frame is much smaller than the
resonant frequency, i.e., $\omega_{\rm
in}\gamma(1-\beta\cos\theta_{\rm in}) \ll \omega_c$.  We
derived (see Appendix) that the electric field ${\bf
E}_{\rm s}(t)$ of the scattered wave is
$$
{\bf E}_{\rm s}(t) = -{r_{\rm c} \over D} E_0 
\frac{\cos\eta \sin\theta_{\rm in} \sin\theta_{\rm out}}
     {\gamma^2(1-\beta\cos\theta_{\rm out})^{2}} e^{i({\bf k} \cdot {\bf D}
  - \omega_{\rm out} t)}\cdot {\bf e}^{\rm out}_1,
\eqno(3)
$$
where ${\bf k}$ the wave-vector, ${\bf D}$ the position
vector from scattering electron to observer, $r_{\rm c}$ the
electron classical radius.  The incident wave has electric
field amplitude $E_0$, and its electric vector has an angle
$\eta$ with respect to ${\bf e}^{\rm in}_1$.  The wave is
scattered to be the out-going wave with only completely
linear polarization, the polarization vector of which (${\bf
e}^{\rm out}_1$) is in the plane of out-gong wave and
magnetic field, in consistent with the results from the QED
calculation shown above. We note that Chou \& Chen (1990)
\markcite{cc90} have calculated Stokes parameters of Thomson
scattered wave in strong magnetic fields and found that the
scattered radiation is always linearly polarized for any
polarized incident wave. Their results serve as an
independent check of the specific case of $\gamma=1$ above.

In 3-D, emission from a single electron is going out in a
micro-cone around the magnetic line, the cross-section profile
is shown in Fig.~2, with the maximum at about
$\theta_{\rm out}={1\over \sqrt{3}\gamma }\simeq$0.6/$\gamma$
(i.e. $0.3^{\circ}$ when $\gamma=100$. See Fig.~2), as deduced from
Eq.~(A9). At this angular radius, the frequency of the scattered
waves should be (from Eq.~2)
$$
 \omega_{\rm out} = 1.5 \;\omega_{\rm in}\gamma^2
 (1-\beta \cos\theta_{\rm in}).
\eqno(2a)
$$
When a line-of-sight goes across the micro-cone, one can
detect an ``S'' shape variation of polarization angle. Note
that this cross section of the micro-cone does not have a
gaussian-shape. All those polarization properties do not vary with
frequency until $\omega_{\rm in}$ is about $10^{14}$ s$^{-1}$.

(2). For resonant scattering at high energy bands (for
$B=10^{12}$G, $\omega_{\rm in} =1.3\;10^{18}$ s$^{-1}$), the
scattered emission ($\omega_{\rm out} =1.7\;10^{21}$
s$^{-1}$) has a filled gaussian-like shape with the total
emission peak in the direction of magnetic field (see Fig.~3),
rather than a micro-cone. Circular polarization is 100\%
near the center, but decreases to zero when the angular
radius $\theta_{\rm out}$ is about $1/\gamma$, and it changes
the sense and gets almost 100\% again when
$\theta_{\rm out}=1.5/\gamma$. In contrast, there is no linear
polarization near the center and the edge, but it peaks up
to almost 100\% when $\theta_{\rm out}=1/\gamma$. The
position angle of linear polarization of the scattered
photon is perpendicular to the co-plane of out-going photon
and the magnetic field, different from the case in radio band.

(3). When $\omega_{\rm in}$ is {\it about} the resonant
frequencies, i.e.  higher than about $10^{14}$ s$^{-1}$ or
lower than $10^{20}$ s$^{-1}$ as seen in Fig.~4, the total
intensity of the scattered photons increases towards the
resonant frequency. However, this increased total intensity
is mainly contributed from circularly polarized emission,
which is shown in the upper plots of Fig.~4 from the fraction
variations of circular polarization with frequency at given
directions of out-going photons.

What we considered in the second and third cases is at
high energy regime with parameters $B=10^{12}$G and $\gamma=100$
in this section.
However, polarization properties of scattered emission we
presented above should be the same for the situation in the
resonant region in the outer magnetosphere of pulsars which
we will not considered in this paper.

Note that all these conclusions for an electron are the same
for a positron because the amplitude of the scattered waves
rests on the incident waves, having nothing to do with the
charge sign of a particle.

\section{The coherent ICS process in pulsar magnetosphere}

Strong polarization is the most outstanding feature of
pulsar emission.  Circular polarization has been detected
from pulsars, and sometimes has sense-reversals across pulse
profile (Han et al. 1998)\markcite{hmxq98}.
In this section, we consider the pulse
polarization in the frame of the ICS model.

As we see from the last section, scattered emission in radio
band from a single electron is completely linearly polarized,
and the position angle follows an ``S'' shape for a sweep of the
line of sight. The scattered waves from electrons in a close
group are coherent, which can produce the circular
polarization detected from pulsars. The amplitude of
coherent emission can be calculated by using
Eq.~(3) for radio emission, so that we can avoid the much
more complicated procedures by using coherent states in
quantum electrodynamics. For example, if two linearly polarized
plane waves {\bf E}$_1$ and {\bf E}$_2$ with same wave vector have
a phase difference $\delta$ and an angle of $\kappa$ between
their polarization vectors, the circular polarization
percentage of the coherently superimposed emission is
$$
{V \over I} = {2E_1E_2\sin\kappa\sin\delta \over E_1^2+E_2^2
+ 2 E_1E_2\cos\kappa\cos\delta}.  \eqno(4a)
$$
For $E_1 = E_2$, Eq.(4a) becomes
$$V/I = \sin\kappa\sin\delta /(1+\cos\kappa\cos\delta). \eqno(4b)$$
This gives considerably circular polarization, as long as the values
of $|\sin\kappa|$ and $|\sin\delta|$ are not too small.  If $\kappa =
\delta = \pi/ 2$, it will be totally circularly polarized.

In the ICS model of radio emission of pulsars, the incident waves
were generated by a sparking due to the breaking down of inner gap.
The waves travel through a vacuum-like pulsar magnetosphere
with a small filling factor of pair plasma 
(i.e., inhomogenous spatially). They encounter a bunch of
relativistic particles that come from another cascade (in the other
side) above the polar cap, and are upscattered coherently (Fig.~5)
to produce the observed radio emission. In a given observational
direction, coherently superpose scattered waves from a bunch of
particles, which is seen from the transient ``mini-beam'' below,
can be almost completely linearly polarized or some times
have very strong circular polarization dependent on geometry.
This is mainly the result of the coherency and the non-random
distributions of the $\delta$ and $\kappa$ in Eq.(4).
However, the observed
integrated pulse profiles of pulsars, is the {\it incoherent} sum
of many samplings of these mini-beams. In the field of view of
a given line of sight, if the sparkings distribute symetrically,
the circular polarizations will cancel each other, and no circular
polarization left finally. Nevertheless, there is remarkable
circular polarization in mean pulses if the sparking distribution
is asymmetric.

In this section, we first describe the coherent ICS
process of a bunch of particles, then we present computer
simulations of their transient beam. Finally we simulate the mean
pulse profiles, which have linear and circular polarization,
S-shaped position angle swing and unpolarized emission as well.

\subsection{The scattered waves from a bunch of particles}

The scattered emission from a single particles is in a micro-cone
of about $0.6/\gamma$ (see Fig.~2 and Sect.~2), hence
only emission from a small area is visible to a given line-of-sight
${\bf n}_0$. Since a particle is moving in a direction
along the magnetic field ${\bf n}_{\rm B}$, the emission
from these particles satisfying $ {\bf n}_B \cdot {\bf n}_0
\sim \cos(0.6/\gamma)$ can be received by an observer.
Using Eq.(2a) (see also Eq.(6) of Qiao \& Lin\markcite{ql98}
1998), combined with Eqs.(11) and (13) of
Qiao \& Lin (1998)\markcite{ql98} which
describe the scattering geometry and how the Lorentz factor of the
particles changes along the field lines, one can find
three emission heights for a given frequency.
At a certain observing frequency, an observer cannot receive
emission at the same time from all zones at three heights produced by
a bunch of particles, but just see a part of one zone
\footnote{
Actually, one just see a ring-like area about
$ {\bf n}_B \cdot {\bf n}_0 \sim \cos(0.6/\gamma)$, where
(1) the scattered emission is at a given observational frequency;
(2) the power reaches its maximum.
}.

The upper-scattered waves from particles in such
a small area can be accumulated coherently for some good reasons.
First of all, the low frequency incident waves that were generated
by one sparking should have a phase coherence when they encounter the
bunch of particles, even if with a small phase dispersion due to
travelling. Secondly, the scattered waves at one given frequency from
these particles in a small visible ring-like area would have a good
phase coherence as well\footnote{
We understand that if this area is much less than the wavelength of
incident waves, there would be coherence. Otherwise, for example,
``mini-beam'' from the outer cone region we discuss below is much larger,
and can't be treated coherently.
}. The complex electric field amplitude contributed from each
particle[see Eq.(3)] reads
$$
{\bf E_s}= C {\cos\eta \sin\theta_{\rm in} \over R}
\exp[{\it i} ({\omega_{\rm in} \over {\it c}}R-{\omega_{\rm out} 
\over {\it c}}{\bf R}  \cdot {\bf n_0}+\phi_0)]
\;  {\bf e}_1^{\rm out} ,
\eqno{(3a)}   
$$ 
here $C$ is a constant for a given $\gamma$.
Vectors {\bf R} = SA and {\bf n$_0$} are
illustrated in Fig.5, and $\phi_0$ is the initial phase of the incident
waves when it was generated by a sparking. Note that $\phi_0$ might be
random for different sparkings. When the incident waves come from one
sparking, we can integrate the complex amplitudes
(rather than the power) from Eq.~(3a)
of scattered waves from all visible particles.
The complex electric field amplitude of the total scattered wave is,
$$
{\bf E} = \int_{ {\bf n}_B \cdot {\bf n}_0 \sim \cos(0.6/\gamma)}
{\bf E}_{\rm s} \; n_{\rm e} \; dV.  \eqno(5)
$$
Here, $n_{\rm e}$ is the number density of particles in the bunch.
The Stokes parameters of this accumulated emission are (Saikia 1988)
\markcite{saik88}$$
\left\{ \begin{array}{lll}
I & = & {1\over 2} [({\bf e}_{10}\cdot {\bf E})^* ({\bf e}_{10}\cdot {\bf 
E}) + ({\bf e}_{20}\cdot {\bf E})^* ({\bf e}_{20}\cdot {\bf E})],\\
Q & = & {1\over 2} [({\bf e}_{10}\cdot {\bf E})^* ({\bf e}_{10}\cdot {\bf 
E}) - ({\bf e}_{20}\cdot {\bf E})^* ({\bf e}_{20}\cdot {\bf E})],\\
U & = & {\rm Re}[({\bf e}_{10}\cdot {\bf E})^* ({\bf e}_{20}\cdot {\bf 
E})],\\
V & = & {\rm Im}[({\bf e}_{10}\cdot {\bf E})^* ({\bf e}_{20}\cdot {\bf E})].
\end{array} \right.
\eqno(6)
$$
Two orthogonal polarization vectors above were defined to be 
$
{\bf e}_{20}  =  {\bf n}_0\times{\bf \Omega}/|{\bf n}_0\times{\bf
\Omega}|,$
and $ {\bf e}_{10}  =  {\bf e}_{20}\times{\bf n}_0$, here $\Omega$
is the direction of rotation axis (Fig.~5).

\subsection{Transient beam from a bunch of particles: simulation}

We now simulate the coherent ICS process of a bunch of particles, and
will present its 2-D emission feature snap-shot.

The particles
produced by the sparkings are moving out along the extremely strong
magnetic fields. The radius of a sparking spot was assumed to have
the same scale as the inner gap height (Gil 1998) \markcite{gil98},
which Deshpande \& Rankin (1999)\markcite{dr99}
estimated to be about 10 meters. We took this
value in the following simulations, but we have found that changing
of this dimensional size does not affect our results.
We also assumed that the number density of particles $n_{\rm e}$ in a bunch
has the maximum in the center and decline towards the edge. We took 
a gaussian distribution for convenience. This number density will be
the natural weight upon the contribution in Eq.(5) from different
visible part of a ring-like area at a given frequency. The wave-phase
of this contribution depends on the location of this visible part
relative to a sparking spot.

The ICS process by one bunch of particles moving along given field lines
forms three ``mini-beams'' of a few degrees in width. They were produced
at emission heights of core and two cones, respectively. 
We assumed that the particles in the bunch have the same Lorentz factor,
and found that the polarization features are quite different between these
mini-beams(see Fig.~6).

As seen in Fig.~6, circular polarization is very strong, even up to
100\%, in the core minibeam, as shown by the open and filled circles.
Emission from the minibeam of inner cone is much
less circularly polarized. However, we could not calculate the beam
feature of the outer cone using coherent superposition method
as in the case of core and inner cone,
since the particles at this larger emission height have much smaller Lorentz
factors, and therefore the visible ring-like area (see Fig.~5) is so large
that the upscattered emission is no longer coherent.

We also made simulations for a bunch of particles with different Lorentz
factors, but the polarization features shown in Fig.6 do not change
significantly. This does not surprise us since the global polarization
pattern is mainly determined by field geometry.

When the line of sight sweeps across a minibeam, we can see a {\it transient}
``subpulse'', as shown in Fig.~7. Note that the width of ``transient''
subpulses is about $0.6^{\circ}$, much larger than $1/\gamma \sim 
0.04^{\circ}$.
When the line of sight sweeps across the 
center of a  core or inner conal minibeam, the circular polarization will
experience a central sense reversal, or else it will be dominated by one
sense, either left hand or right hand according to its traversal relative
to the minibeam. 

The position angles at a given longitude of {\it transient} ``subpulses''
have diverse
values around the projection of the magnetic field (see Fig.6 and
Fig.7). The variation range of position angles is larger for core
emission, but smaller for conal beam. When many such ``subpulses'' from 
one minibeam is summed up, the mean position angle at the given longitude
will be averaged to be the central value, which is determined by the
projection of magnetic field lines. Therefore in our model, the mean
position angle of pulsar emission beam naturally reflects the
geometry of magnetic field lines. As we will show in our simulation the
position angle of mean pulse profiles should have ``S'' shapes (see section 
3.3).

Note that these transient ``subpulses'' are not the subpulse
we observe from pulsars. The time scale for the transient beam\footnote{This
corresponds to the time scale of the inner gap sparking} to
exist is very short,
about $10^{-5}$~s. A real observation sample with
extremely high time resolution (e.g. a few $\mu$s) will detect one point
from only one transient ``subpulse''. In this case, a strong (even near 100\%)
circular polarization could be observed. Observational evidence was
presented by Cognard et al. (1996)\markcite{cstt96}.
However, if sampling time is longer (e.g. ms), what one receive is the
incoherent summation of emission from hundreds of transient beams at a
given longitude. This largely diminishes the circular polarization,
but does not affect much on the linear one.

\subsection{Polarization features of integrated pulse profiles}

Integrated profiles of pulsars are characterized by linear polarization,
circular polarization, the sweep of position angle, and substantial
amount of unpolarized emission. Our simulation shows these characters
can all be reproduced. In our model, at every longitude, an observer
can see the scattered emission from particles in the
many sparking-produced bunches around his line of sight. To get the
integrated profile, we sum up incoherently
the Stokes parameters of the scattered waves from all these bunches.
For the simplicity of simulation, we assumed the particles in one bunch have
the same Lorentz factor, since changing the Lorentz factor or considering
the energy distribution of these particles should not alter the results
following.

We used the geometry presented in Fig.~5 for our model calculation,
which helped to determine the different $\theta_{\rm in}$ values
for a sparking to every bunch in the ICS process.
The key factor is the distribution of sparking spots in the polar
cap. In the inner gap, a sparking is triggered when a $\gamma$-photon
encounters considerable $B_\bot$, namely the perpendicular component
of magnetic field, and produces an $e^{\pm}$ pair. A pair production
cascade forms and the inner gap is discharged. This seems more likely to
occur in a peripheral parts of the polar cap due to large $B_\bot$
while the central parts are inconducive to discharge. But at the
edge of the gap, there is null electric field and thus it is
impossible for a sparking to occur (Ruderman \& Suthland 1975)\markcite{rs75}.
Therefore, the preferred region of gap discharge is located between
the center and the edge on the polar cap. We assume in our simulation
that the sparking spots have a gaussian distribution radially
from the magnetic axis, with a maximum at
$\theta \sim 0.75\theta_{\rm pc}$, where $\theta_{\rm
pc}$ is the radius of the polar cap. However, changing the form of
this radial distribution will not affect our results qualitatively.

In the inner gap model, the electric potential across the gap is produced
via monopolar generation. When the magnetic axis inclines from the
rotation axis, the potential which accelerates the particles are
different in the part of gap nearer to the rotation axis compared
to that in the further part. Therefore, the probability of polar gap
discharging varies with the distance to the rotation axis. This distribution
asymmetry with respect to the magnetic axis has been included in our
simulations. The probability of the sparkings was assummed
to decrease exponentially with the azimuthal angle from the projection
of the rotational axis. It is several times larger in the nearest part 
to the rotation axis than that in the farthest part.

In Fig.8, we present a set of polarization profiles with various
impact angle $\beta$. We found that the integrated profiles are
highly linear polarized in general, and their polarization angles
follow nice S-shapes. One can immediately see that the position
angle curves have a larger ``maximum sweep rate'' for a smaller
$\beta$, in excellent accordance with the rotating vector model.

One important result is the antisymmetric circular polarization
found in the central or core component, which can be up to 20\%.
As we mentioned before, circular polarization can be up to 100\%
on some part of ``subpulses'', but depolarization occurs in the
process of incoherent summation of circular polarization of opposite
senses. This also produces a substantial amount of unpolarized emission.
For the linearly polarized emission, position angles of transient
``subpulses'' vary, causing further depolarization. 

However, in the conal components, circular polarization is
insignificant, mainly because it is originally weak in the minibeam
(see Sect.3.2). Since there is negligible variation of position angles
of ``subpulse'', almost no linear depolarization presents, and therefore,
the conal components are always highly linearly polarized.

\section{Discussion and Conclusions}

Qiao (1988) and \markcite{qiao88}
\markcite{ql98}
Qiao \& Lin (1998) proposed an inverse Compton scattering model for
radio pulsar emission. The simulations presented in this paper show
that linear as well as circular polarization can be obtained from coherent
inverse Compton scattering process from a bunch of particles, although
the radio radiation of a single electron is highly linearly polarized.
In the simulation process, there are two assumptions we have made,
namely the coherence of the outgoing waves and small velocity dispersion
of particles in a bunch.
The circular polarization is favorable to be produced in emission
regions near the polar cap in the ICS model. Observational data show
that the `core' emission is radiated at a place relatively close to
the surface of the neutron star (Rankin 1993)\markcite{ran93},
the `inner cone'
is emitted in a higher region, and the `outer cone' in the highest.
Our simulations show that circular polarization tends to appear in
the core components, while the `outer cone' emission is linearly
polarized.

One might have noticed that the circular polarization from our simulation
of the ICS model is always strong and antisymmetric in the central part
of integrated pulsar profiles. It is of intrinsic origin. 
However, almost no circular polarization appears in the conal components
according to the ICS model, therefore the observed one must be converted
from the strong linear polarization in propagation process (Radhakrishnan
\& Rankin 1990; Han et al. 1998; von Hoensbroech \& Lesch 1999).
\markcite{rr90}\markcite{hmxq98}\markcite{vl99}

One important ingredient to produce the circular polarization in the
ICS model is the non-radial asymmetry of sparking distribution. We found
from our summation that, without the non-radial asymmetry, circular
polarization will be canceled by incoherent summation. The same conclusion
has been reached by Radhakrishnan \& Rankin (1990)\markcite{rr90}
when they studied circular polarization using curvature radiation.

The polarization of the individual pulses or micro-pulses, however,
should be related to  the time resolution of observations according to
the ICS model. With high time resolution (e.g. short to $\sim10\mu$s),
each sample will take emission from few transient beam that has strong
linear and circular polarization. Depolarization happens during the
sample time of a low resolution. This can serve as a test for our model.

We emphasize that the width of observed subpulses is not necessary
associated with $1/\gamma$ suggested by Radhakrishnan \& Rankin
(1990)\markcite{rr90}.
The width of real subpulses observed from pulsars is generally 2-3 degrees,
which was used to estimate $\gamma$ to be less than several tens.
In our simulation, the width of {\it transient} ``subpulse'' produced
by a bunch of particles along certain field lines reflect the width
of the minibeam, which is mainly determined by the dimension of the bunch.
The bunch radius near the star surface was taken to be 10 meters, 
corresponding to about one degree in longitude for core emission,
and a little bit larger for cones. The width of a real observed sub-pulse
might be related to the dimension of magnetic tube for the bunches
more often to flow-out. Therefore, the upper limit of $\gamma$ can be
released to $10^3$ or even higher.

In our simulation, the low-frequency waves are assumed to be
monochromatic. We have also made simulations for various
$\omega_{\rm in}$, and found that the polarization properties of
transient beam and integrated profiles are quite similar to the
results we presented above.

\acknowledgments
We are very grateful to Dr. Bing Zhang and B.H. Hong for many extensive
discussions, to Drs. R.N. Manchester, Q.H. Luo, Bing Zhang and anonymous
referee for very helpful comments and suggestions.
This work is supported by National Nature Sciences Foundation of China,
by the Climbing project of China, and
by the Youth Foundation of Peking University. JLH acknowledges for
the financial support from the Education Ministry of China and the
Su-Shu Huang Astrophysics Research Foundation of the Chinese Academy
of Sciences.

\section*{Appendix: The amplitude of scattered wave in strong magnetic field}

We consider the complex amplitude of scattered wave in the
electron-rest-frame, and then make Lorentz transformation
for our case.  The equation of motion for an electron with
$\gamma=1$ in a magnetic field $B$ is
$$
m\stackrel{..}{\bf r}=e{\bf E}(t)+{e \over c}\stackrel{.}{\bf r} \times
	{\bf B}.	\eqno(A.1)
$$
Here, $e$ is the electron charge, and $c$ is the speed of light.
For conditions $|\dot{\bf r}| \ll c$ and $\omega_{\rm in} \ll \omega_c$, 
the magnetic component of the incident wave and the radiation damping force
have been neglected. The electric field of the incident wave is taken as
${\bf E}(t)
= {\bf E}_0 \exp({-i\omega_{\rm in} t}) = {\bf e}^{\rm in}\;
 E_0 \exp({-i\omega_{\rm in} t})$ (here ${\bf e}^{\rm in}$ is the polarization
vector for incident wave) and the position
vector of the electron as ${\bf r}(t)  =  {\bf r}_0
	\exp({-i\omega_{\rm in} t})$.
Generally, ${\bf E}_0$ and ${\bf r}_0$ are complex vectors. One then can get
$$
m\omega_{\rm in}^2{\bf r}_0 +¡¡e{\bf E}_0
	- i{e \omega_{\rm in} \over c}{\bf r}_0\times{\bf B}=0. \eqno(A.2)
$$

For the incident wave in $x$-$z$ plane (see Fig.1), fundamental linear
polarization vectors of incident wave were defined
${\bf e}^{\rm in}_1 =  \{ -\cos\theta_{\rm in},\; 0,\; \sin\theta_{\rm in}\}$
and ${\bf e}^{\rm in}_2  =  \{ 0,\; -1,\; 0\}$. The magnetic field
${\bf B}=\{0,\; 0,\; B\}$. The fundamental vectors of scattered waves were
chosen to be $ {\bf e}^{\rm out}_1  =  \{ -\cos\theta_{\rm out}\cos\phi,\;
-\cos\theta_{\rm out}\sin\phi,\; \sin\theta_{\rm out} \}$, and 
${\bf e}^{\rm out}_2  =  \{\sin\phi,\; -\cos\phi,\; 0 \}$. See Fig.~1
for the parameters of geometry.  For a given incident wave,
$ {\bf e}^{\rm in} = \cos\eta{\bf e}^{\rm in}_1 + \sin\eta{\bf e}^{\rm in}_2$.
Here ${\bf e}^{\rm in}$ is a complex vector, and hence $\eta$ is a complex
number generally. One can get the solutin of Eq.(A1) as:
$$
{\bf r}_0 = {\nu r_c \over 1-\mu^2} \left(
\begin{array}{l}
 \cos\eta\cos\theta_{\rm in} + i\mu\sin\eta \\
 \sin\eta - i\mu\cos\eta\cos\theta_{\rm in}\\
 \cos\eta\sin\theta_{\rm in}(\mu^2-1) 
\end{array} 
\right),  \eqno(A.3)
$$
here $\mu \equiv \omega_c / \omega_{\rm in}$,  
     $\nu \equiv E_0 c^2 /(e \omega_{\rm in}^2)$,
and  $ r_c ={e^2 \over mc^2}$. 
In case of strong magnetic field, $\mu \gg 1$, and
if $\eta \neq \pi / 2$, Eq.(A.3) can be simplified 
as being
$$
{\bf r}_0 = \{0,\; 0,\; -\nu r_c\cos\eta\sin\theta\}.	\eqno(A.4)
$$
The electric field for the scattered wave can be written as
(for $|\dot{\bf r}| \ll c$)
$$
\begin{array}{lll}
{\bf E}_{\rm s}(t) & = &
 {e \over cD} [{\bf n}\times({\bf n}\times\stackrel{..}{\bf r})]_{\rm ret}\\
             & = &
 -{r_{\rm c} \over D} E_0\cos\eta\sin\theta_{\rm in}\sin\theta_{\rm out}\; 
\exp [i{\omega_{\rm in} \over c} ({\bf n} \cdot {\bf D} - c t)]
\; {\bf e}^{\rm out}_1 .
\end{array}	\eqno(A.5)
$$
Here, 
${\bf n} = \{ \sin\theta_{\rm out}\cos\phi,\; 
           \sin\theta_{\rm out}\sin\phi,\;
           \cos\theta_{\rm out} \}$
and ${\bf D}$ is the position vector from a scattering
point to an observer, $D={|{\bf D}|}$.  From Eq.(A.5) one see that
the scattered wave has only the term of ${\bf e}^{\rm out}_1$, and there
is no ${\bf e}^{\rm out}_2$ term, which implies the completely linear
polarization of the scattered waves. The differential cross section is
$$
d\sigma_\eta = {E_s^2 D^2 d\Omega_{\rm out} \over E_0^2}
= r_c^2\cos^2\eta\sin^2\theta_{\rm in}\sin^2\theta_{\rm out}d\Omega,
\eqno(A.6)
$$
here $d\Omega_{\rm out}$ is the solid angle of the scattered wave.
The total cross section is 
$$
\sigma_\eta = \int d\sigma_\eta =
\sigma_{\small \rm Th} \cos^2\eta\sin^2\theta_{\rm in},
\eqno(A.7)
$$
here, $\sigma_{\small \rm Th} = {8\pi\over 3}r_c^2$ is the Thomson section.
For $\eta = 0$ (i.e. just for one polarization of ${\bf e}^{\rm in}_1$),
one can get $\sigma|_{\eta=0} = \sigma_{\small \rm Th} \sin^2\theta_{\rm 
in}$,
which is consistent with the Eq.(16) at the lower frequency
limit ($\omega_{\rm c}\gg \omega_{\rm in}$) in Herold's (1979) when
${\omega\over \omega_{\rm B}\pm\omega}=0$.

In the extreme case of $\eta=\pi/2$, the value of $|{\bf
r}_0|$ is extremely small (with a factor of $(\omega_{\rm
in} / \omega_{\rm c})^2$) for the scattered waves in the
radio band, so we will not discuss it in following.

For an electron with $\gamma>1$, one can easily found through
Lorentz transformation that the polarization vector of a
wave is the same in an alternative inertial frame, but the
complex amplitude of electric field of the wave in moving
frame is $\gamma (1-\beta\cos\theta_{\rm in})$ times
that in a rest frame. In the electron rest frame (the
frame moving with a electron with Lorentz factor $\gamma$),
the complex amplitude of the electric field for the incident
wave is
$ E_0^{\small R} = \gamma (1-\beta\cos\theta_{\rm in})E_0$.
The complex amplitude for the scattered wave in the electron rest frame is
$ E_{\rm s}^{\small R} = -{r_{\rm c} \over D} E_0 \cos\eta\sin\theta_{\rm in}
\sin\theta_{\rm out}/  [\gamma(1- \beta\cos\theta_{\rm out})]$,
hence, in the laboratory frame it becomes
$$
E_{\rm s}  =  \gamma(1+\beta\cos\theta_{\rm out}^{\small R})E_s^{\small R} \\
 =  -{r_{\rm c} \over D} E_0 \cos\eta\sin\theta_{\rm in}\sin\theta_{\rm out}
[\gamma(1-\beta\cos\theta_{\rm out})]^{-2}.
\eqno(A.8)
$$
Therefore, for a moving electron with Lorentz factor $\gamma$, the electric
field of the scattered wave becomes
$$
{\bf E}_s(t) =
 -{r_{\rm c} \over D} E_0\cos\eta\sin\theta_{\rm in}\sin\theta_{\rm out}\;
 [\gamma(1-\beta\cos\theta_{\rm out})]^{-2}\;
\exp [i{\omega_{\rm in} \over c} ({\bf n} \cdot {\bf D} - c t)]
\;  {\bf e}^{\rm out}_1 .
\eqno(A.9)
$$

\newpage

\begin{figure}			
\psfig{file=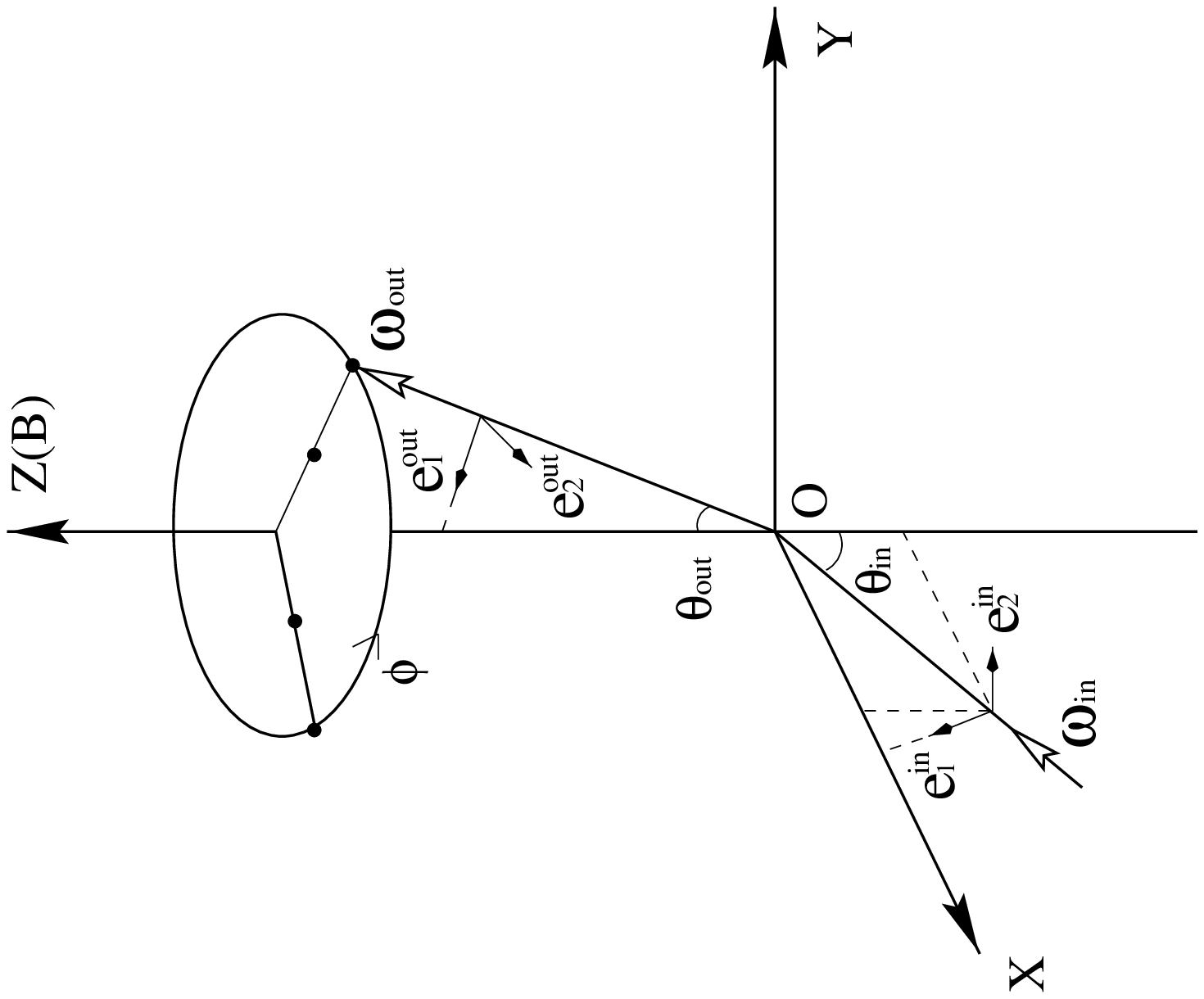,height=88mm,width=100mm,angle=270}
\caption{The geometry of the ICS process for a single
electron.  The incident photon is coming as $\omega_{\rm
in}$ in the x-z plane, and the scattered photons are going
out around the magnetic field line (i.e., the $z$ axis). The
geometrical parameters used in the rest of the paper have
been marked. In x-z plane, $\phi=0$.
}
\end{figure}
\begin{figure}		
\psfig{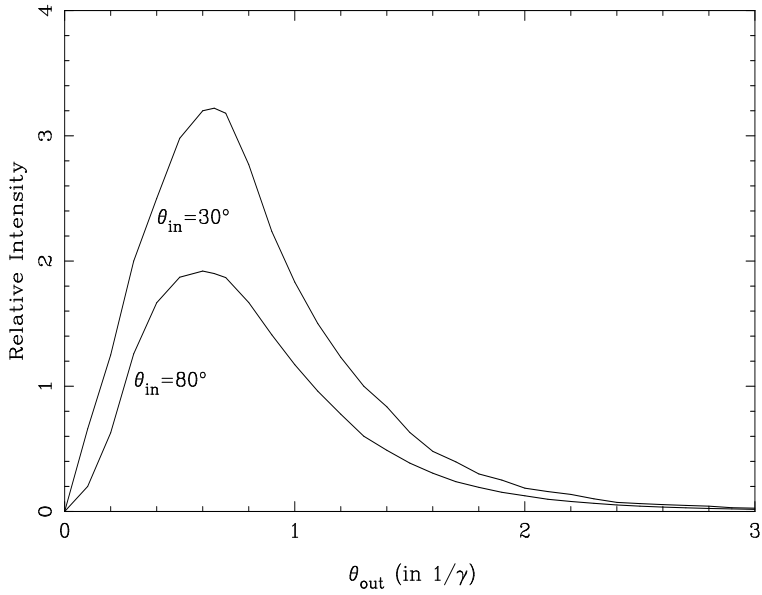}
\caption{The total and linearly polarized intensities vary
with the distance from the field direction, forming an
micro-cone of out-going photons. The curves were
calculated for $\theta_{\rm in} = 30^o$ and $80^o$.  Other
parameters were taken as: $ B = 0.1 B_q$ (here $\omega_c =
7.75\; 10^{19}$ s$^{-1}$), $\gamma=100$, $\omega_{\rm
in}=10^5$ s$^{-1}$.
}
\end{figure}
\begin{figure}		
\psfig{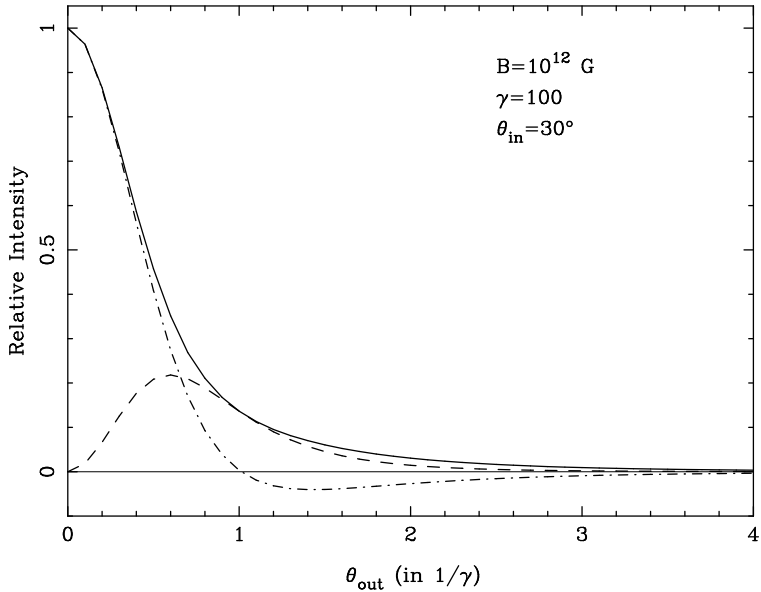}
\caption{The resonantly scattered waves: the total, linear
and circular polarization intensity varying with the distance
from the magnetic field direction were plotted as a solid line,
dashed line and dash-dot-dash line, respectively.
\label{fig3}}
\end{figure}
\begin{figure}		
\psfig{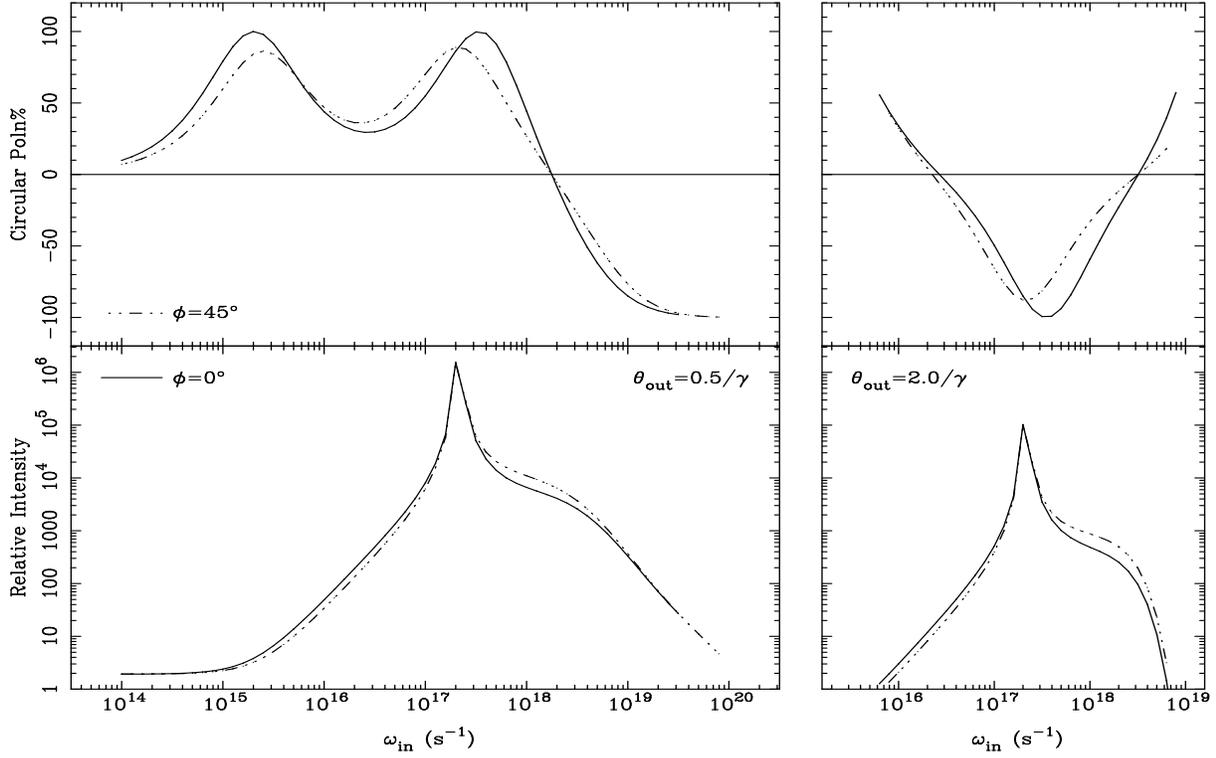}
\caption{The frequency dependence of total intensity
(lower plots) and circular polarization percentage
(upper plots) of the scattered waves {\it
about} the resonant frequency, in four given directions of
the out-going photon beam as
$(\phi, \theta_{\rm out}) = (0^{\circ}, 0.5/\gamma)$,
$(45^{\circ}, 0.5/\gamma)$,
$(0^{\circ}, 2.0/\gamma)$, and
$(45^{\circ}, 2.0/\gamma)$. See Fig.~1 for geometry. The solid
lines are plotted for the azimuthal angle $\phi=0^{\circ}$, and
dash-dot-dot-dot lines for $\phi=45^{\circ}$. The left is plotted
for emission at the angular radii $\theta_{\rm
out}=0.5/\gamma$, and the right for $\theta_{\rm
out}=2/\gamma$.
\label{fig4}}
\end{figure}
\begin{figure}		
\psfig{file=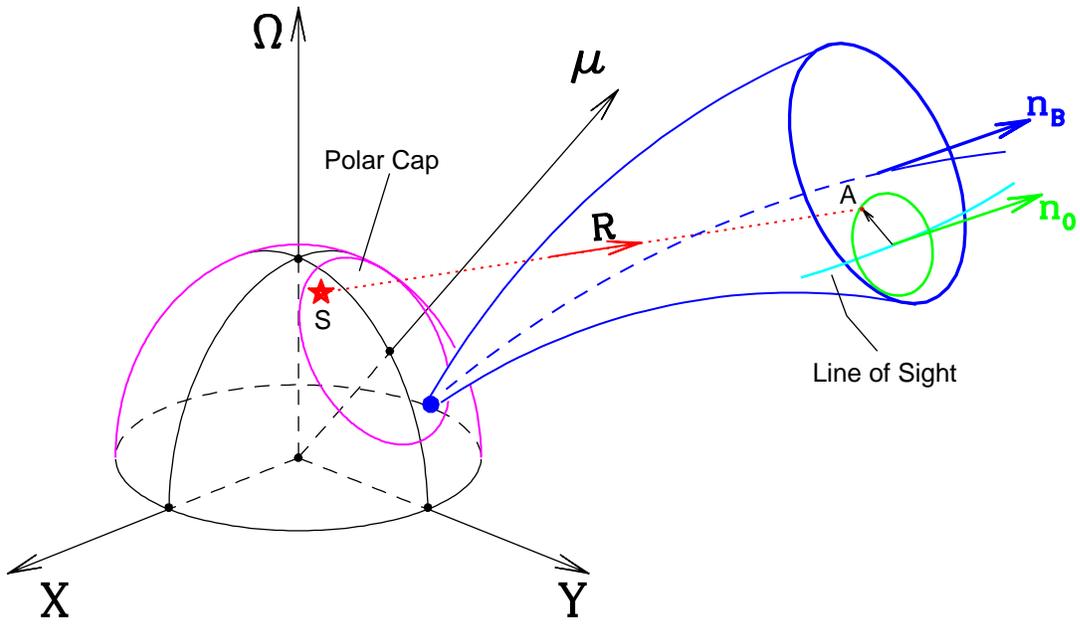,height=150mm,width=150mm,angle=0}
\caption{The geometry of the coherent ICS process. The sparking occurs
at a point S above the polar cap, and produces the incident waves that
are scattered by a bunch of particles.  The emission visible to an
observer at {\bf n}$_0$ is from a ring-like area satisfying ${\bf n}_B
\cdot {\bf n}_0 \sim \cos(0.6/\gamma)$. Point $A$ is on this ring.} 
\end{figure}
\begin{figure}		
\centerline{\psfig{file=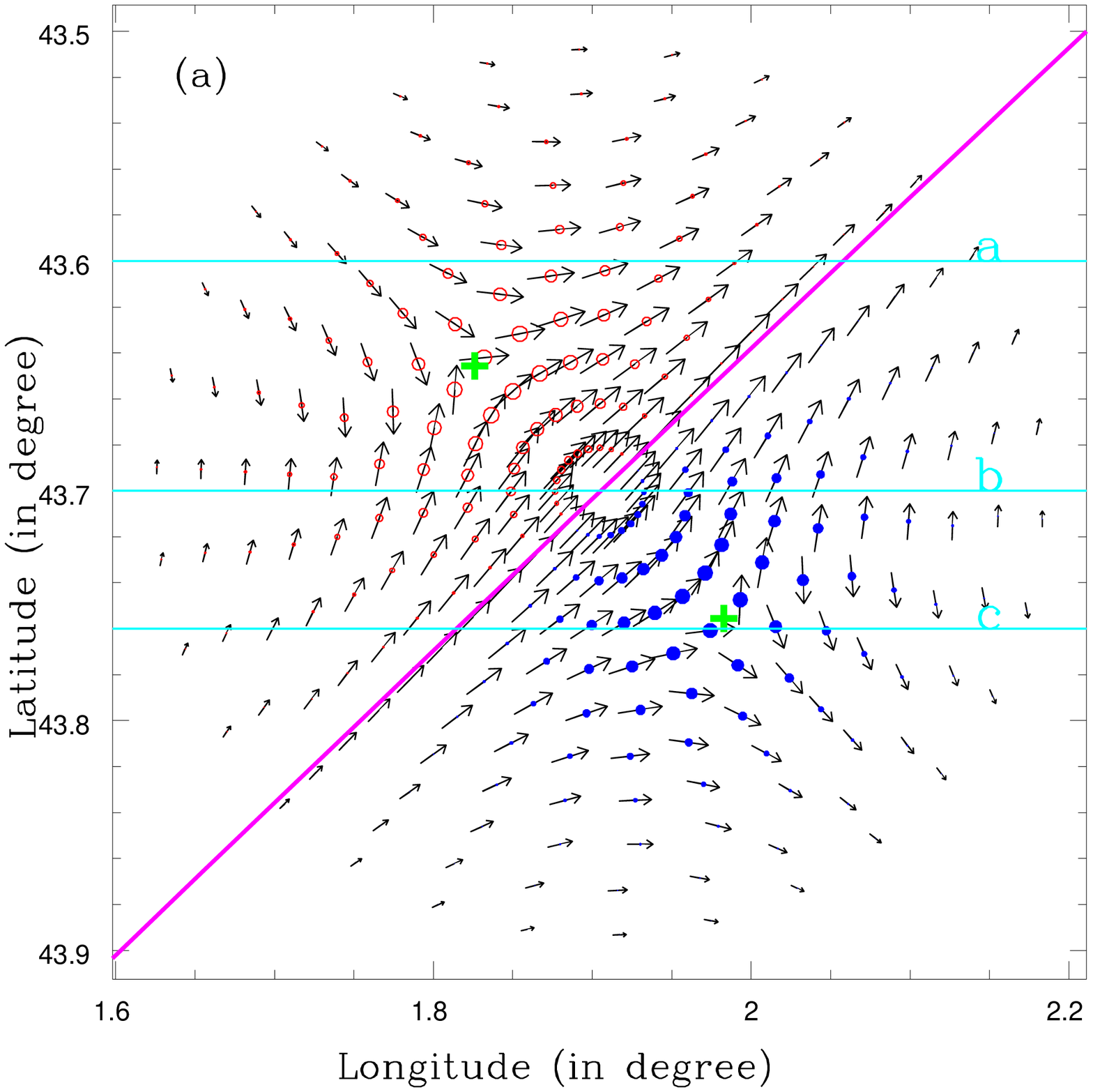,height=80mm,width=80mm,angle=0}
\psfig{file=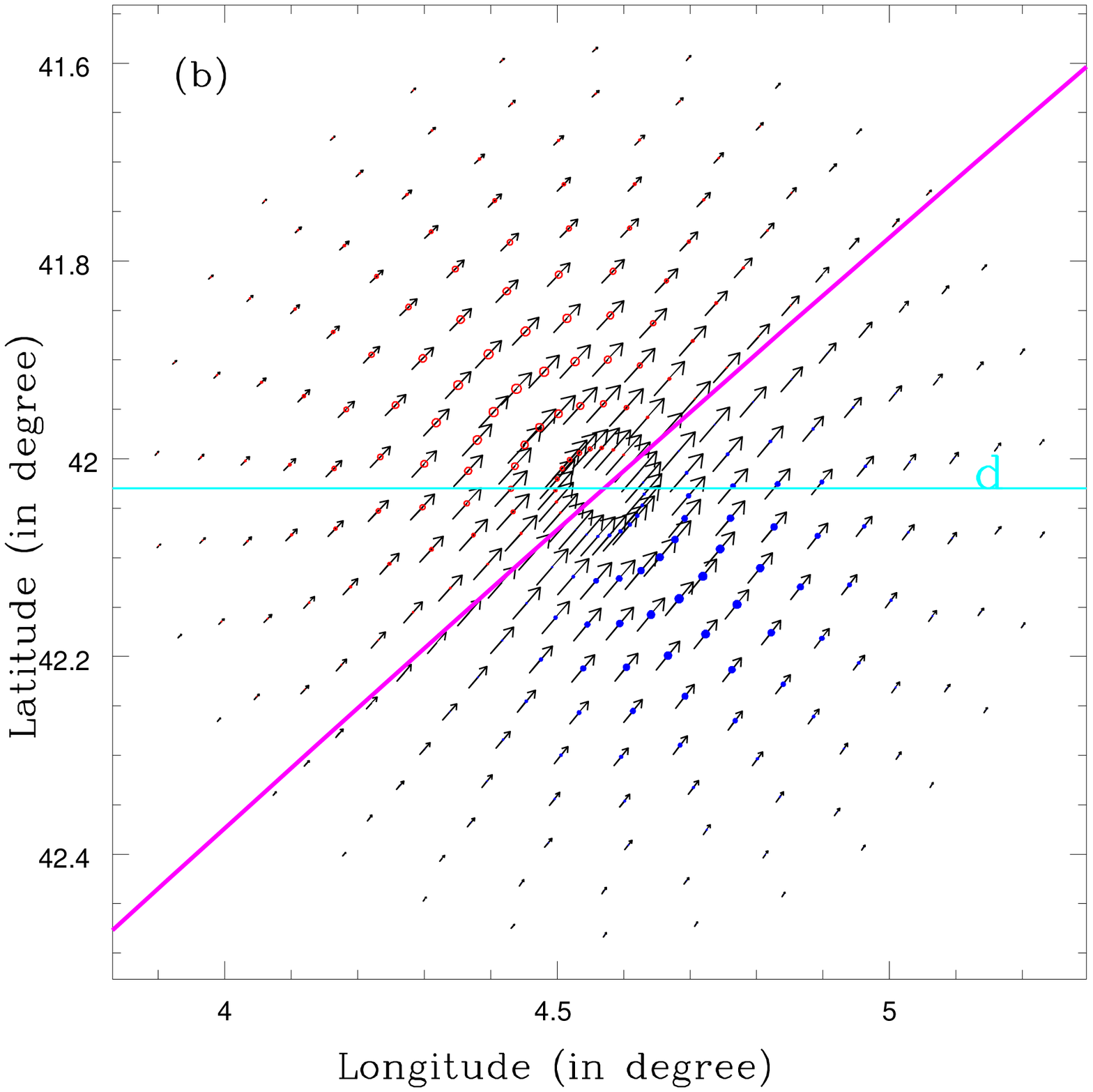,height=80mm,width=80mm,angle=0}}
\caption{Polarization features of transient {\bf (a)} core and {\bf (b)}
inner conal minibeams. The vectors stand for the polarization vector,
their length is proportional to intensity.
The filled and open circles represent the circular polarization of
two senses, their size is proportional to circular polarization
percentage.
The circular polarization at two points in the core minibeam
marked by the crosses gets to the maximum of 100\%, with
position angle jumps and null linear polarization.
The sizes of circle in Fig.6b have been amplified by
4 times for clarity.
The thick line is the projection of the curved magnetic field.
The line of sight goes across the beam, for example, along lines
$a$, $b$ and $c$ in Fig.~6a and $d$ in Fig.~6b.
The radius of bunch is $R_{\rm b}=10$ m, magnetic inclination angle
$\alpha=45^\circ$, the bunch location $\phi_{\rm c}=225^\circ$, and
$\gamma_0=2000$, $\omega_{\rm out}=1$ GHz.
}
\end{figure}
\begin{figure}		
\centerline{\psfig{file=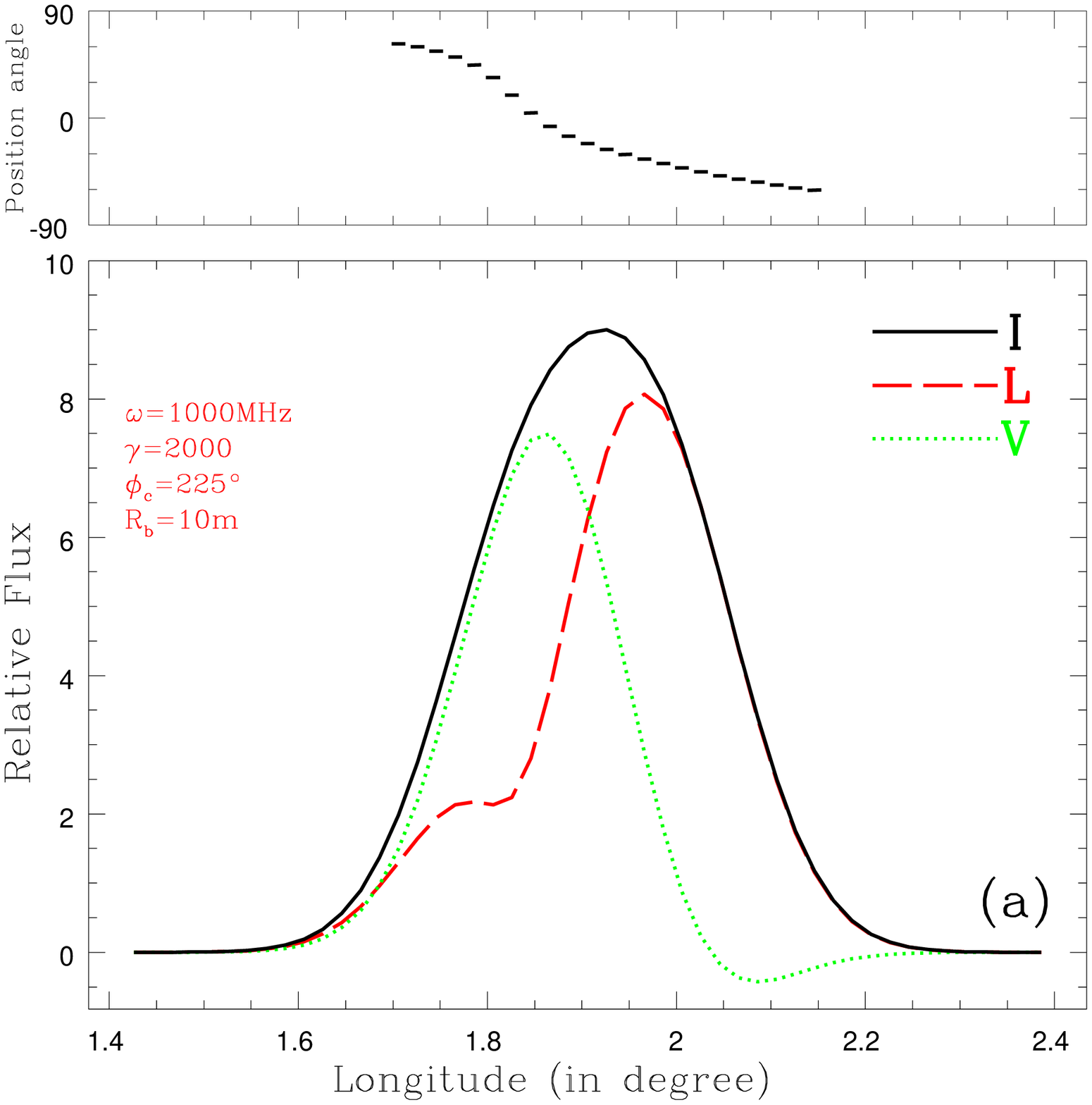,height=80mm,width=80mm,angle=0}
\psfig{file=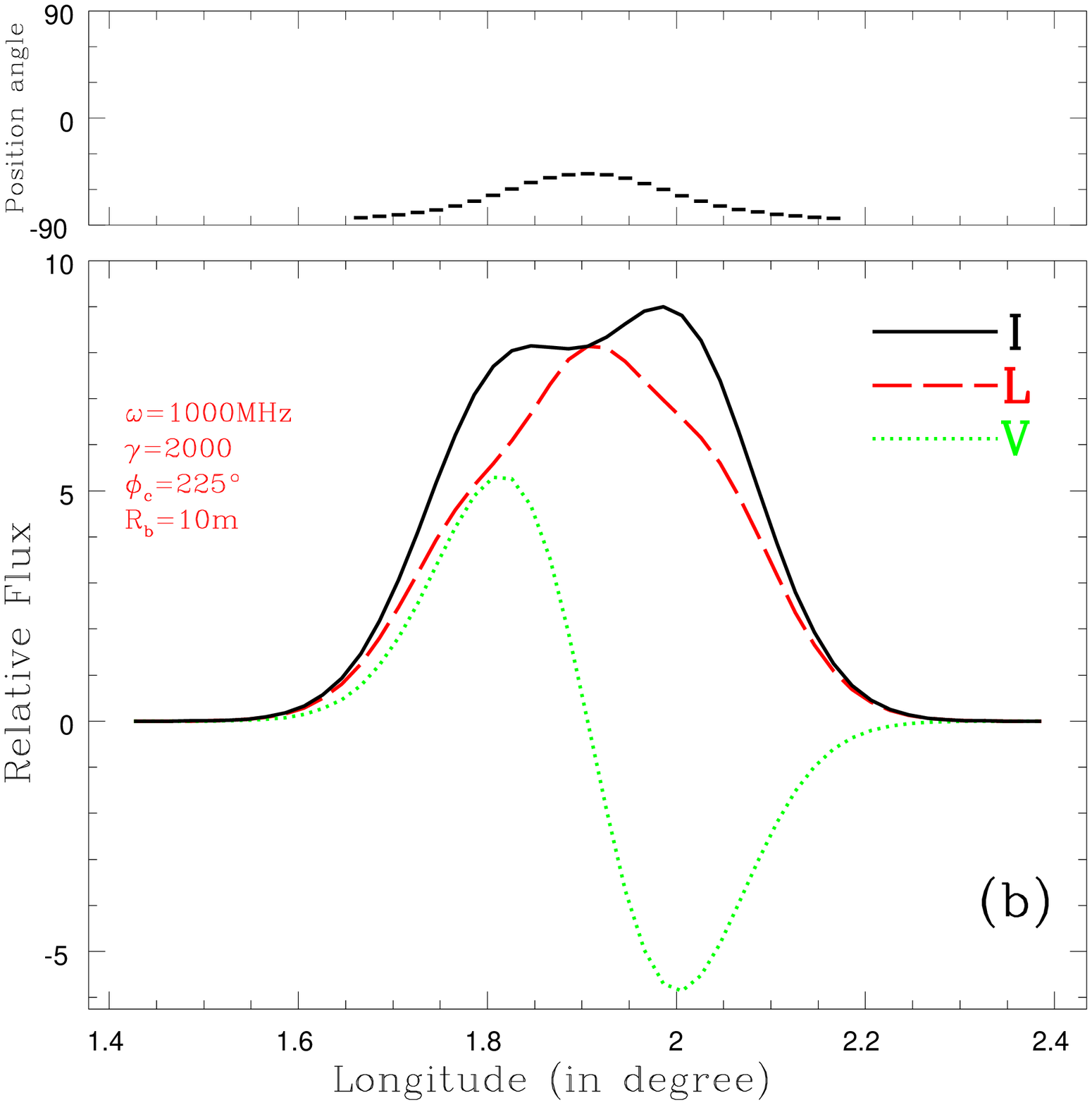,height=80mm,width=80mm,angle=0}}
\centerline{\psfig{file=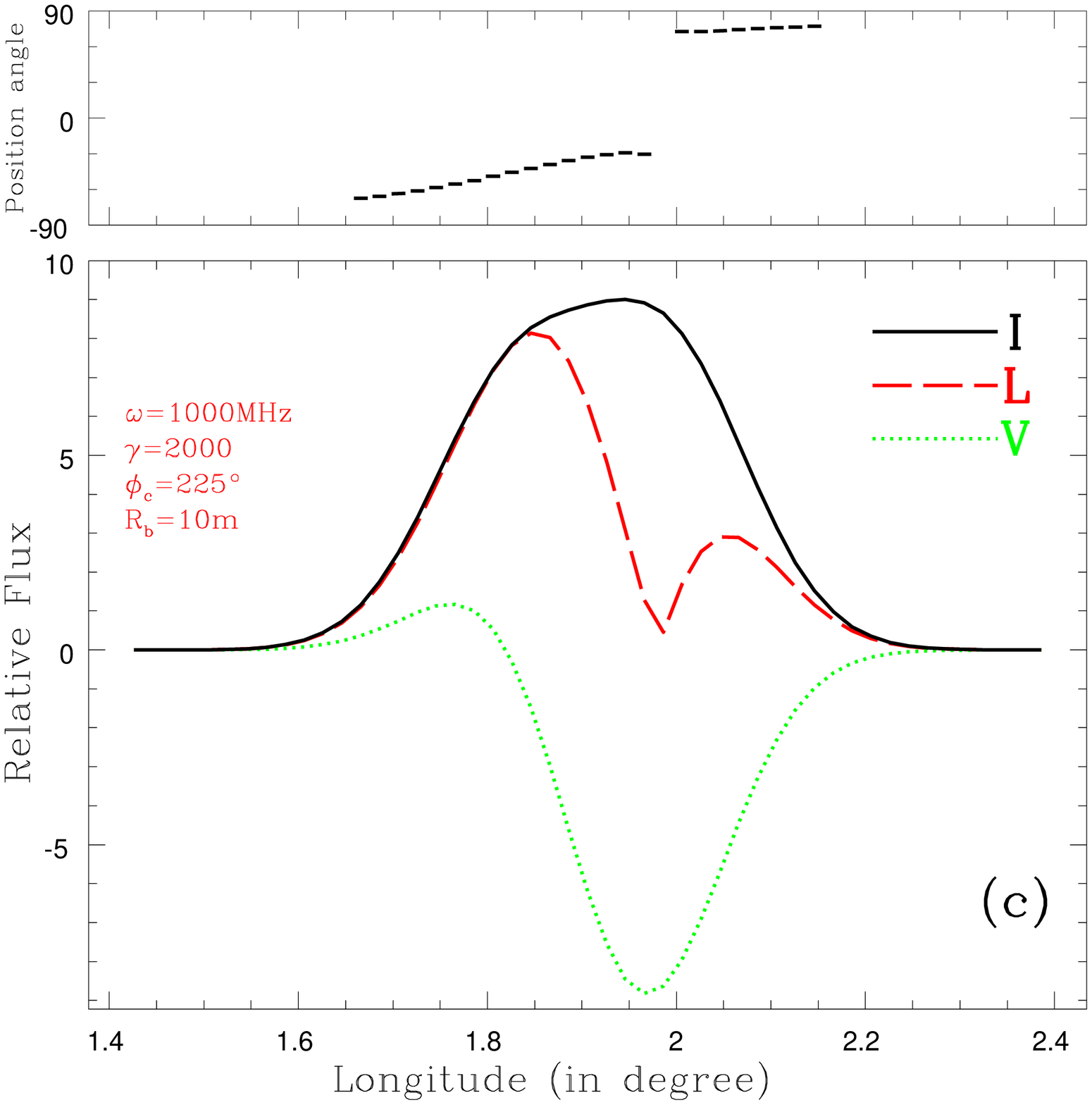,height=80mm,width=80mm,angle=0}
\psfig{file=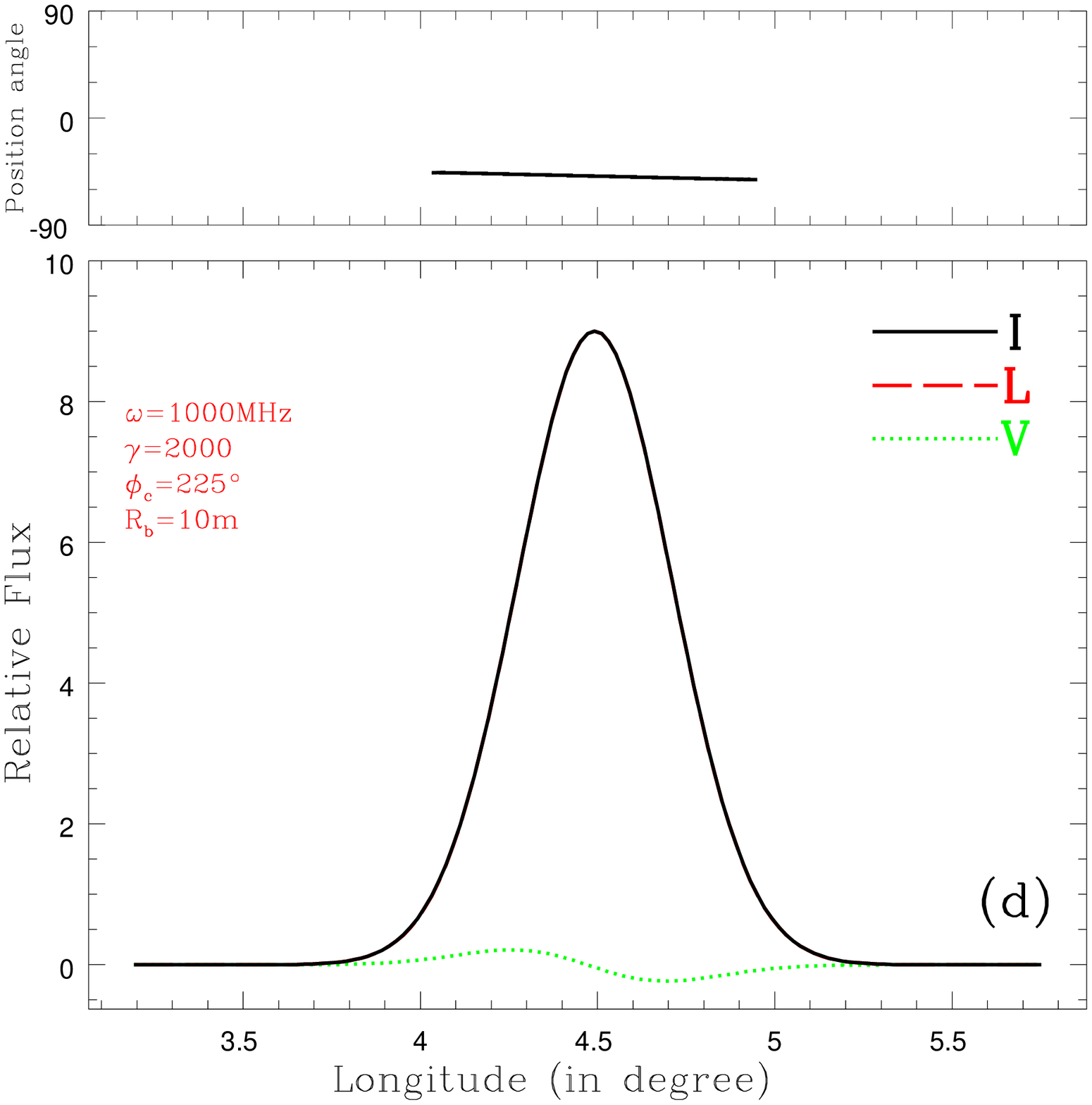,height=80mm,width=80mm,angle=0}}
\caption{Cross-sections of transient beams, cut by the lines of
sight $a$, $b$, $c$ and $d$ in Fig~6.
The obvious features of these ``subpulses'' are the presence of
different circular polarization patterns and the variation of position
angles. The circular polarization percentage and position angle variation
range  of inner cone emission (d) both are much less than that of core
emission (a, b, c).  
}
\end{figure}
%
\begin{figure}  
\centerline{\psfig{figure=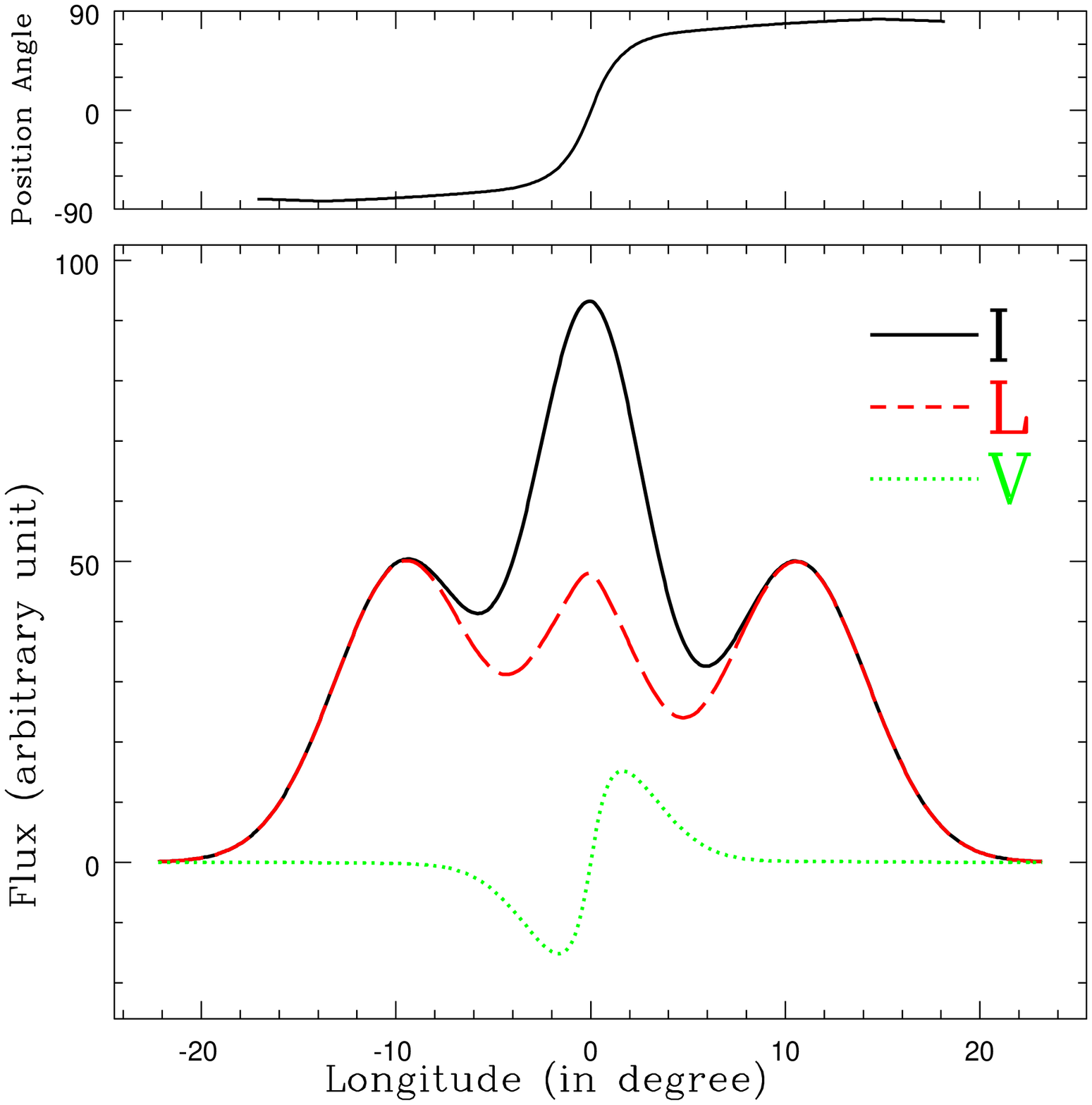,angle=0,height=6cm,width=6cm}\psfig{figure=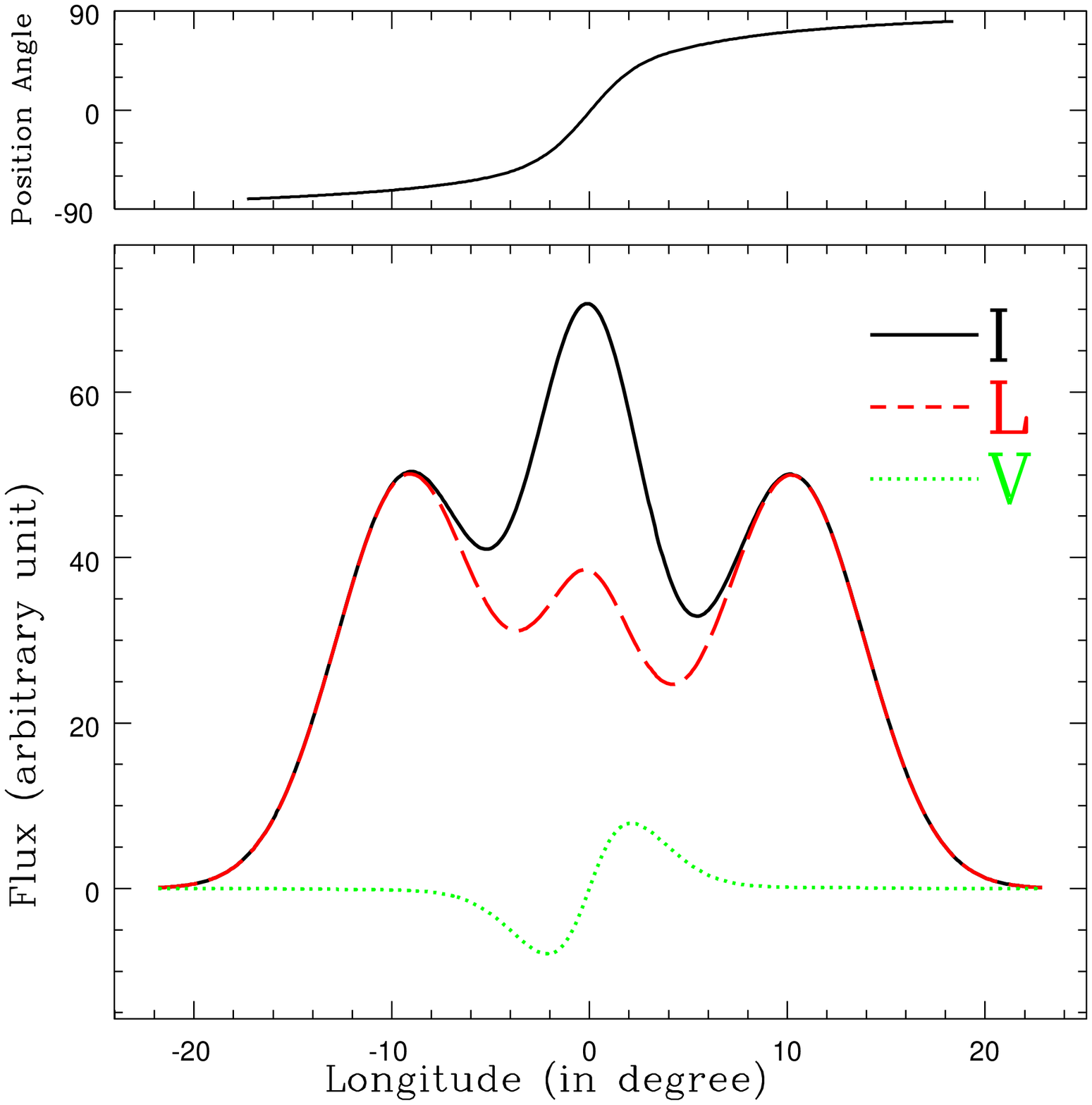,angle=0,height=6cm,width=6cm}\psfig{figure=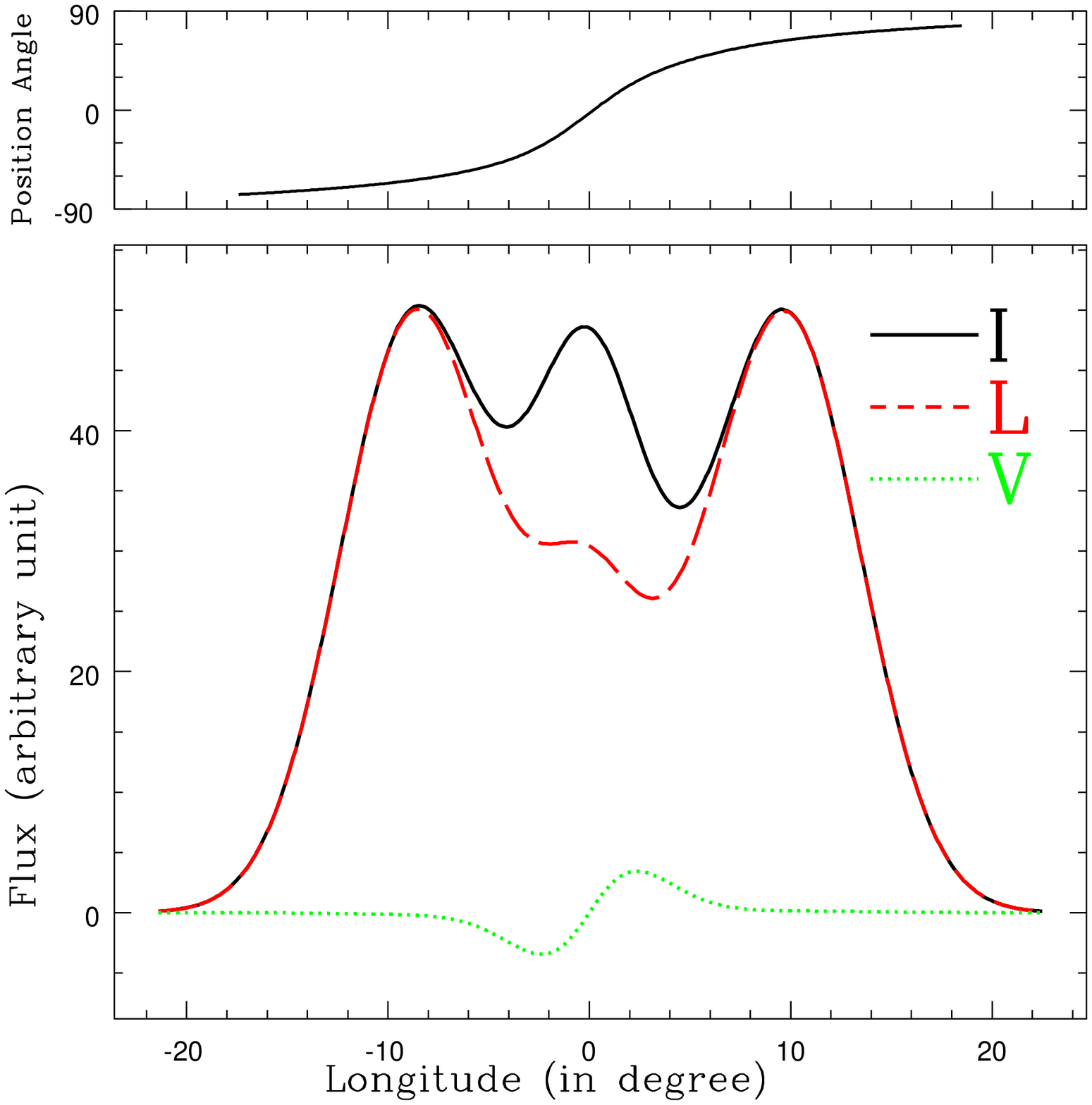,angle=0,height=6cm,width=6cm}}
\centerline{\psfig{figure=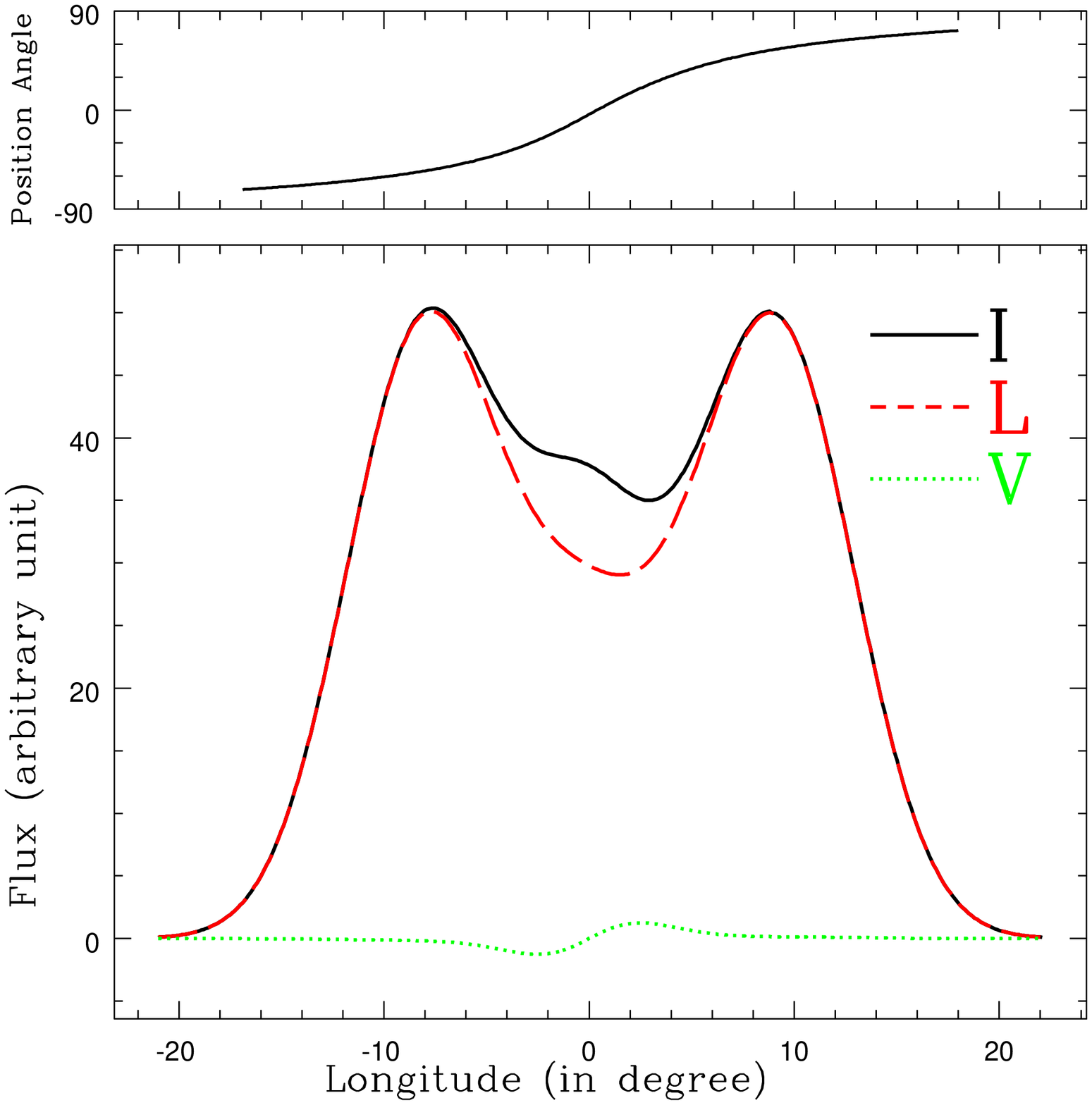,angle=0,height=6cm,width=6cm}\psfig{figure=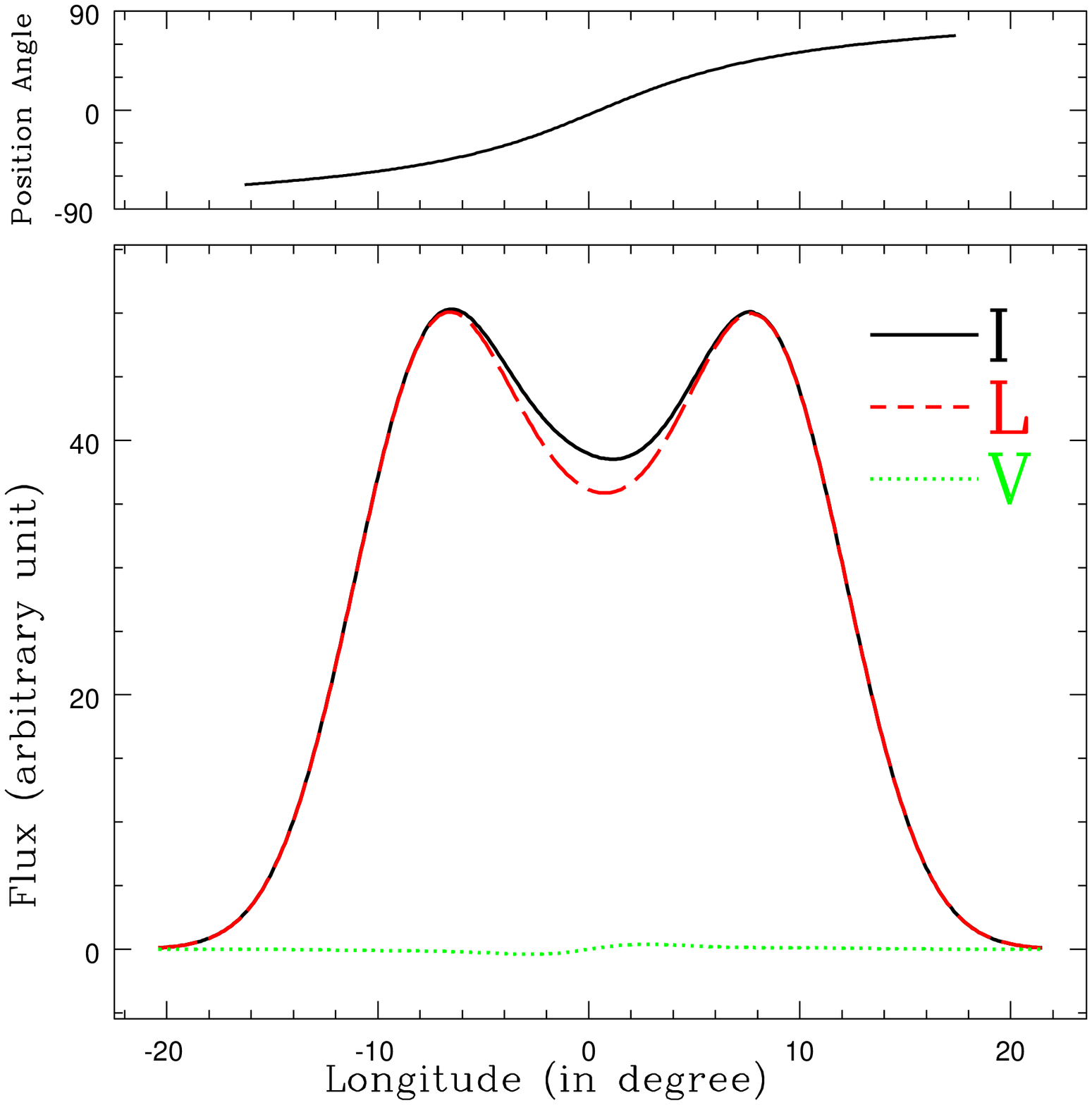,angle=0,height=6cm,width=6cm}\psfig{figure=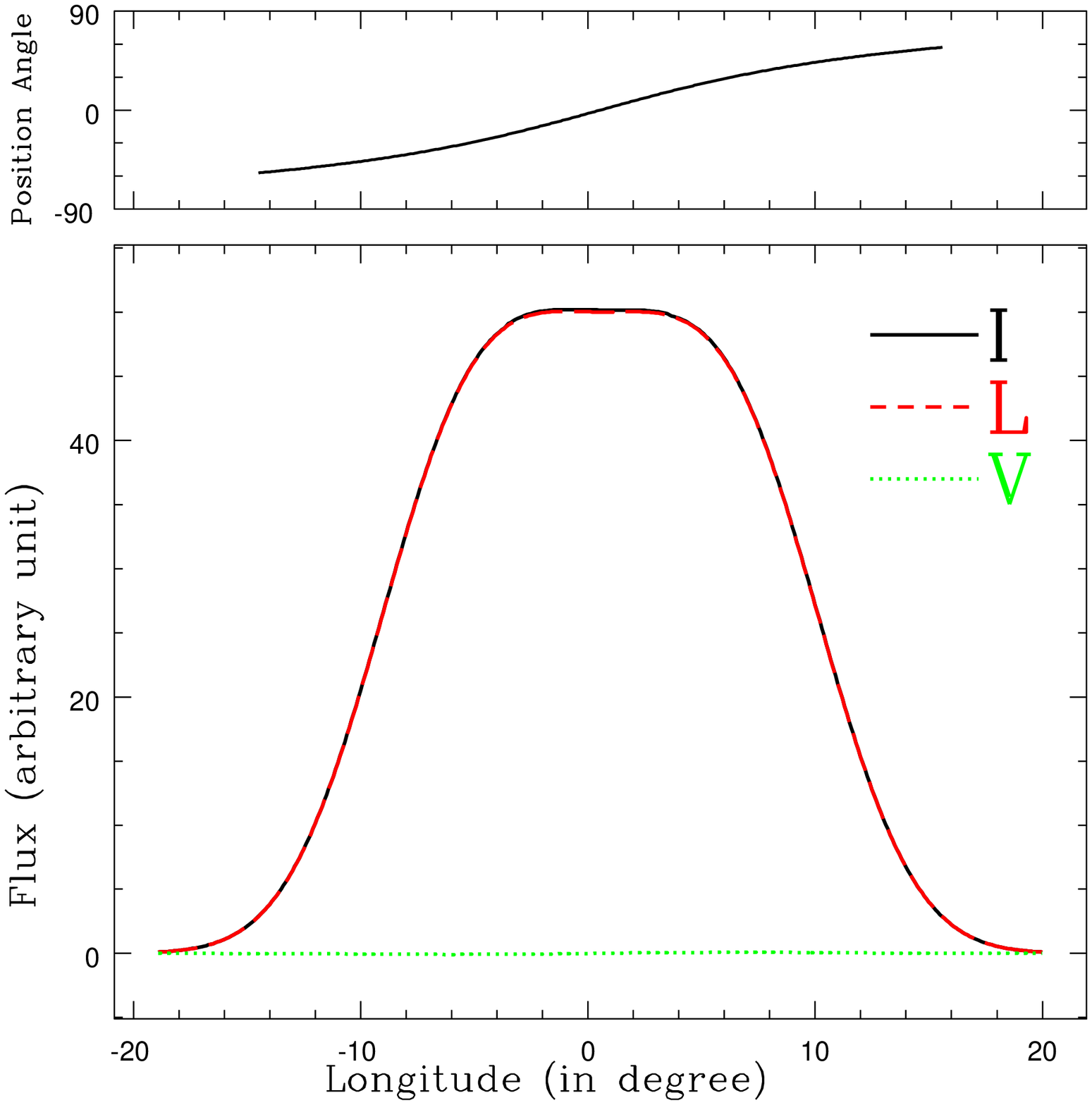,angle=0,height=6cm,width=6cm}}
\caption[]{A set of integrated pulse profiles, simulated for the
normalized impact angle $\beta_{\rm n} = 0.15$, 0.30, 0.45, 0.60, 0.75, and 0.9.
An S-shaped position angle swing, linear polarization, an antisymmetric
type of net circular polarization and substantial amount of unpolarized
emission can be presented from the ICS model. A small
beam-shift between core and conal components due to different emission
heights has been included in the simulation.
}
\end{figure}
\end{document}